\documentstyle[preprint,aps]{revtex}

\preprint{\bf PREPRINT}

\def\h{{\bf h}}
\def\z{{\bf{\hat z}}}
\def\r{{\bf r}}
\def\be{\begin{equation}}
\def\ee{\end{equation}}

\begin{document}
\columnsep0.1truecm
\draft
\title{The effect of field tilting on the dynamics of vortices 
pinned by correlated disorder}
\author{ N. Shnerb}
\address{ Lyman Laboratory  of Physics, Harvard  University,
Cambridge, MA 02135}
\date{\today}
\maketitle

\begin{abstract}
Low-temperature dynamics of flux lines  in high temperature, type II,
 superconductors in the presence of correlated disorder in the form of 
columnar defects   is discussed.  
The effect of tilting the applied magnetic field with respect to the 
column's directions is considered,  using  the non-Hermitian quantum
 mechanics technique evaluated  by Hatano and Nelson \cite {hat}. 
It is shown  that the critical current, as well as the vortex transport 
properties below this current, may be determined by ``surface excitations'', 
i.e., by the roughness of the flux line near the edges of the sample, which 
dominated  the bulk jumps. Phase space considerations determine the critical
 thickness of the sample, below which the tilt induced surface transport
 exceeds the bulk mechanism. This critical length, which depend on the
 tilt angle as well as the directions of the perpendicular field and 
the suppercurrent, diverge at the delocalization transition.
\end{abstract}
\vspace{0.25in} 
\vskip.5pc

   Flux lines response functions in cuperate high-temperature superconductors 
have attracted considerable  interest in recent years \cite {RMP}. 
In order to avoid dissipation of energy as a result of flux lines 
motion driven by the 
superconducting current, these lines should be pinned by crystal impurities
\cite {anderson}. It turned out \cite{nel-vin}
 that the pinning  is much  stronger (especially when many 
vortex interaction is taken into account)
when these impurities are in form of  correlated  disorder (such as twin 
boundaries  \cite{experim1} or  columnar defects \cite{experim2}) compared
to the case of point disorder, resulting from vacancies of Oxygen atoms
\cite {point}. However,  Nelson and Vinokur \cite {nel-vin} have pointed out
that the correlated defects pinning become less effective in cases where the 
direction of the external magnetic field is tilted with respect to the 
defects, which we take to be along the ${\bf{\hat z}}$ direction.  
At some critical tilt, for which the energy per unit length of the defect 
is less than the energy associated with the perpendicular field, a 
pinning-depinning   phase transition occur and 
the flux lines delocalized.

 Critical bulk current and vortex dynamics below this current for flux lines 
in the presence of columnar defects  have been 
considered in \cite {nel-vin}. The authors, using the mapping of 
flux lines in $d+1$ dimensional superconductor to the world lines of bosons
in a $d-$dimensional quantum system, identified  the phase space diagram of 
the system which contain a high temperature ``superfluid'' and low-temperature
``bose glass'' phases, as well as Mott insulator at the matching field, 
$B_\phi = n_{pin} \phi_0 $, for which there is one flux line per defect. 
At low temperature, this matching  field separates the ``dilute'' 
region of the bose glass
phase, for which the vortex lines are pinned individually by the defects,
(i.e., $a_0 \sim (\phi_0/B)^{1/2}$, the Abrikosov lattice constant, is much 
bigger than $d$, the typical distance between two columnar defects) 
from the high density region, for which interactions are important in 
determining the localization length and transport properties of the flux 
lines.

 In the low field region, the  vortices are localized by the interaction with
the correlated defects. Each defect is the analog of a $2D$ potential well
which we ??? take as a cylindrical square well such as $V(r)  = -U_0$ for 
$r<b_0$ and $V(r) = 0$ for $r>b_0$. The temperature of the superconductor, in
turn, corresponds to the Plank constant $\hbar$ of the quantum boson problem. 
For the dilute vortex arrays, where the pinning energy is larger then the interaction energy, there are two regimes. For low temperatures ($T<<T^*$ , 
$T^* \equiv \sqrt{U_0 \epsilon_1 }b_0$) the localization length $l_\perp$ 
is on the radius of the defect; i.e., of order $b_0$; so that each flux line 
is localized by $\it{ one}$ defect. On the other hand, for $T>> T^*$, the localization length of one defect grows exponentially with $T^2$, and the flux line is then localized be several defects, forming an effective $d$ dimensional potential well in the corresponding boson system.

 The response of the flux line to superconducting current in the plane 
perpendicular to the vortex direction ${\bf J \perp B}$ translates itself 
into the response of the boson system to an applied electric field. 
For vortices oriented in the ${\bf \hat z}$ direction, the Lorentz force 
per unit length of the vortex is given by

\be
{\bf f_L} = {\phi_0 \over c} {\bf \hat z} \times {\bf J}
\ee

Which is the analog of a boson with charge $\phi_0$ interacting with the electric field ${\bf E} = {1 \over c}  {\bf \hat z} \times {\bf J}$.

   Above the critical current $J_c$, the vortices are no more localized ant there is no superconductivity (in the sense of dissipation free current) anymore. Below this critical current, the mechanism for flux flow is tunneling via thermally activated ``half loops'' (or, in the boson dynamics, tunneling into the conduction bend). For currents smaller than $j_1$, the half loop transverse displacement exceeds the mean distance between occupied pinning sites, and for thick sample the flux lines move via the nucleation of superkinks, the analog of the Mott variable range hopping (VRH) in doped semiconductors \cite{8}.

   The depinning  of the flux line as a result of external field  tilt 
has been carefully investigated by Hatano and Nelson \cite {hat}. The   
Hemiltonian of the corresponding boson problem is no longer Harmitian;
the kinetic term of the Hamiltonian ${\bf p}^2 / (2 \epsilon_1)$ (${\bf p}
\equiv -iT \nabla$)  is subject to imaginary gauge transformation and takes 
the form $({\bf p}+i{\bf h})^2/(2\epsilon_1)$, where $\h$ is related to the 
perpendicular magnetic field ${\bf H}_\perp$ via $\h = \phi_0 {\bf H}_\perp
/(4 \pi)$. As a result,    
there are two solutions for each  localized (real  energy spectrum) state,
corresponding to the eigenstates of the Hamiltonian and its complex conjugate.
These solutions, termed $\psi_R$ and $\psi_L$, correspond to the right or left
``tilting'' of the localized solution of the untilted Hamiltonian, i.e.,
\be
\psi_{R,L} ({\bf r}) = \exp(\pm {\bf h \cdot r})  \psi({\bf r}).
\ee 
 The  probability distribution   to find the flux line 
at the point {\bf r}  at a distance $\tau$ from  the surface of the sample
is given by
\begin{eqnarray}
P({\bf r},\tau) = {\cal Z}^{-1} <\psi^f|\exp(-(L-\tau)H/T)|{\bf r} > 
<{\bf r}|  \exp(-\tau H/T)| \psi^i>\\
= {\cal Z}^{-1} \sum_{m,n}  <\psi^f|m_L><m_R|{\bf r}><{\bf r}|n_L>
<n_R|\psi^i> e^{-(\tau E_m + (L-\tau)E_n)/T}
\end{eqnarray}  
such that, as $L \to \infty$, the probability distribution of the flux line at
 the surface is proportional to $<{\bf r}|gs,{L,R}> = \psi^{L,R}_gs({\bf r})$ 
where $|gs_{L,R}>$  are the left and the right ground state, respectively. Deep
 in the bulk, the distribution is given by $P({\bf r}, L/2) = \psi_gs^R 
\psi_{gs}^L = \psi_{untilted}^2$, i.e., in the localized regime, the flux line
changes its shape near the surface, while remains uneffected in the bulk (see
Fig. 1).  Typically, the ``surface roughness associated with the tilt extends into the bulk up to characteristic distance $\tau^*$, which diverges as the tilting angle approaches the critical angle, for which the flux line delocalizes and the current response become linear.

 In this communication, we study the  the effect of the tilt on  the 
   
flux lines response  to superconducting  currents 
in the regime  where the tilting angle is smaller then critical, i.e., in 
the bose glass phase where the  vortex lines are localized. We assume that the
thickness of the sample is large enough, such that it is much larger than 
the dimension of the optimal excitation along the $\z$ exis, Moreover, we 
address  only the dilute limit, for which the transverse ($xy$) displacement 
of these excitations is less than $a_0$, so that the interaction is taken
 into account by filling up the localized states in order of increasing
 energy up  to the  chemical potential $\mu$.

 Let us consider first the critical currents.  This  current 
is determined by the binding free energy $U(T)$ as well as the 
localization length $l_\perp = 1/\kappa $. Modeling the defect as a square 
well in the boson system,    $\kappa$ is related to $U(T)$ by 
$\kappa = \sqrt{2 U(T) \epsilon_1}/T$. Of these two,  $l_\perp $ is changed as 
the magnetic field is tilted. Near the surface of the sample, the 
localization length should be 
$l_\perp(h,\theta) = 1/(\kappa - h | \cos(\theta)|)$,
where $\theta$ is the angle between the Lorentz force $f_L$ 
(perpendicular to the supercurrent  ${\bf J}$, which we take to be in the 
$xy$ plane)  and the tilting field $\h$. The absolute value is needed  
for the case of $|\theta|>{\pi over 2}$,
for which the critical current is dominated by the ``tail'' of the flux line 
on the other end of the sample, as shown in Fig. 1. Thus, the critical 
current will   take the form,
\be
J_c (T,h,\theta)  = {c U(T) (\kappa(T) - h |\cos(\theta)|) \over \phi_0 }.
\ee
This critical current is determined by the surface ends of the vortex, 
for which the effect of the tilt is maximal. However, the ``creep'' of the 
vortex in the direction of the tilting field is limited by the effect of ``image vortices'' which should be introduced in order to satisfy the boundary condition on the surface \cite{nel-vin}. Thus, for very small perpendicular magnetic field, where the surface roughness extension $\tau^*$ is less then the London length, the tilt have no effect on the vortex pinning and there is no change 
in the response functions of the flux system.

For currents less then  critical, the thermally assisted 
flux flow (TAFF) theory  of the vortex transport gives   the resistivity 
$\rho = {\cal E}/J$ as
\be 
\rho = \rho_0 e^{\delta F /T}
\ee
where $\delta F$ is the energy barrier for flux line jumps. Our basic 
observation is that  deep in the bulk there is 
no influence of the tilt, so that the energy barriers for nucleating  half loops
or double kinks are the same. The physical reason for it  is that, although 
the perpendicular field {\it decreases} the energy barrier for one side of the 
kink / loop, it {\it increases} the energy  needed for the other side. 
The main effect of the  tilt comes from 
{\it surface  kinks / loops}, 
for which the energy barrier really decreases. Although the resulting 
energy barrier $\delta F$ is smaller then the bulk one, so that the 
``resistivity'' associated with it is exponentially smaller, one should take 
into account the phase space prefactor of these two mechanisms - the number 
of surface kinks available is determined by the surface roughness, i.e., by
$\tau^*$, while the number of bulk kinks is of order $(L-\tau^*)/Z$, where 
$Z$ is the distance for which the half loop / kink extends along the relevant 
defect. Thus the nature of the current response is determined by the 
thickness of the sample - for $L>L_c(\h)$, the bulk excitations will 
dominate and the response is tilt independent, while for $L<L_c$ 
surface excitations become important and the voltage drop will be 
tilt dependent. As $h \to h_c$ (where $h_c$ is
the critical field above which the flux lines delocalize) $L_c$ diverges, 
 so that near the depinning transition the resistivity of the system goes 
continuesly to zero.

 In order to estimate the relevant quantities in the tilted case, we use the 
expression for the free energy of the surface excitations in the presence of 
the tilt. Consider now surface excitation of the flux line with line tension 
$\epsilon_1$ which extends for a distance $z$ along the pin and has 
perpendicular extent $r$. The free energy of such jump is given by

\be
\delta F = {\epsilon_1 r^2 \over z} + U_0 z - f_L rz - hr \cos(\theta)
\ee

for the ``half loop'' surface excitations. if the jump is due to the 
nucleation  of superkinks, one should take into account the energy difference 
between different rods at distance $r$. This, in turn, is determined by the 
density of states at the chemical potential $g(\mu)$ \cite{nel-vin}. and the 
free energy is,

\be
\delta F = {\epsilon_1 r^2 \over z} +{ z \over g(\mu) r^2} - f_L rz - 
hr \cos(\theta)  
\ee

The resulting saddle point free energies are
\be
\delta F^* = (E_k - h |\cos(\theta)| d) (J_1/J) \qquad {\rm half  loops}
\ee

\be
\delta F^* = (E_k - h |\cos(\theta)| d) (J_0/J)^{1/3} \qquad {\rm superkinks}
\ee
where $E_k = \sqrt{\epsilon_1 U_0 }d$, $J_1 = cU_0/(\phi_0 d)$ and 
$J_0 = c/(\phi_0 g(\mu) d^3)$, for $d$ the average spacing between unoccupied 
pins.

Let us estimate now the phase space for such surface excitations, i.e., the 
width  of the region in which the roughness takes place. Using Eq. (3) one 
finds that the crossover between the surface ($P(\r) \sim \psi^{R,L}(\r)$)
and the bulk, for which $P(\r)$ is the same for the tilted and the untilted
situation, is determined by the quantity

\be
Y(\r) = \sum_m <m_L|\r> \exp(-\tau E_m /T)
\ee
where $\tau$ is the distance from the surface. The transition to the surface 
behavior takes place when $Y(\r)$ becomes $\r$ independent, and thus absorbed 
into the normalization factor for $P(\r)$. Typically, this happens when (10) 
is not  dominated by the $\tau$ dependent exponential factor, since then the 
summation over $m$ is determined by the delocalized states, yielding an $\r$
independent result. Thus, for the case of half loop tunneling, the width of 
the surface roughness will be $\tau^*(h,\theta) \approx T/E^*(h,\theta)$ where
$T \kappa(E^*) = h$. This gives us the estimate

\be
\tau^*_{loops} = {\epsilon_1 \over T(h_c^2 - h^2)}.
\ee   

For the superkinks tunneling, the energy $E_m$may be given by 
$1/(g(\mu)r_m^2)$, so that the energy exponent become negligible as $\tau << 
\tau^*$, where \cite{hat},

\be
\tau^* = {T^3 g(\mu) \over (h_c - h)^2}
\ee

 The phase space  of the surface excitations is given by the width of the 
surface region divided by the ``width'' of the typical excitation 
$z^*_{surface}$.
Using the above expressions for the free energy of the kinks/loops, one finds 
that 

\be
z^*_{surface,loops} (J,h,\theta) = 
{c (\sqrt{\epsilon_1 U_0} - 
h|\cos(\theta)|) \over J \phi_0}
\ee
and
\be
z^*_{surface,kinks} (J,h,\theta) = {c (E_k - 
h|\cos(\theta)|) \over J \phi_0 d}
\ee

The resulting resistivity in thick samples will be determined by adding in 
parallel the $\tau^*/z^*$ ``surface resistors'' with $\rho = e^{\delta 
F^*_{surface}/T}$ to the system of $(L-\tau^*)/Z^*_{bulk}(J)$ ``bulk 
resistors'' with $Z^*_{bulk} = z^*_{surface}(h=0)$. While the surface 
roughness does not depend on the angle between the current and the 
transverse magnetic field, the width of the jump, as well as the free 
energy barrier, do depend on it. It turns out that the resistivity in the 
``perpendicular''  direction (${\bf f_L \perp \h \perp \z}$) is independent 
of the tilt. For other directions of the superconducting current, there will 
be a crossover length $L_c$ below which the surface loops dominate the jumps. 
For any tilt less then critical, the surface roughness is finite, so that as 
$L \to 
\infty$ bulk excitations are clearly the preferred jumping mechanism, but as
 $h \to h_c$, the width of the surface roughness become comparable with the 
sample thickness for each finite sample and one sees a crossover to surface
 excitation dominated transport. The critical length is related to the 
parameters above as 

\be 
L_c(J,h,\theta) = \tau^*({Z^*_{bulk} \over z^*_{surface}} 
\exp(\delta F^*_{bulk} -\delta F^*_{surface}) +1 ).
\ee
 
There are two reasons for the divergence of $L_c$ as $h \to h_c$; one is the 
divergence of $\tau^*$, the other is the fact that $z_{surface}^* \to 0$, 
yielding infinite phase space for the surface excitations. However, there is
 a limitation on the minimal width $z$ of the jumps; as $z \ to \lambda$, the 
London length, 
self interaction of the flux line lock the kink/loop, so that $\lambda$ 
sets the minimal excitation extent along the $\z$ axis. For the region in 
parameter space for which $z^* >> \lambda$, the critical length will grow like 
$(h|\cos(\theta)| - h_c)^{-2}$for loop transport, and as 
$(h|\cos(\theta)| - h_c)^{-3}$ for kinks. On the other hand, as one approach 
the critical tilt, the region 
\be
(h_c - h|\cos(\theta)|) < {\lambda J \over c \phi_0}
\ee
is entered, in which the excitations width could not shrink anymore.
In that case the critical length diverges like  
$(h|\cos(\theta)| - h_c)^{-1}$ for loops, and as $(h|\cos(\theta)| - h_c)^{-2}$
for kinks.

\section*{Acknowledgments}
I would like to thank D.R. Nelson and N. Hatano for most helpful discussions 
and comments. This research is supported by the Rothschild foundation.

-----------------------

{\normalsize\bf Figure Captions}

Figure 1:{Flux lines locaqlized by columnar defects in the presence of 
perpendicular field ${\bf H_\perp}$. The surface roughness extends distance 
$\tau$ into the bulk, and the Lorentz force $f_L$, is at angle $\theta$ to 
the tilting field. For $|\theta| > \pi/2$, the contribution of the surface 
roughness to the vortex transport comes from the ``tail'' at the lower end 
of the sample in the figure.}

\end{document}